\documentclass[a4paper,twocolumn,prl]{revtex4}
\usepackage{graphicx}          
\usepackage{dcolumn}
\usepackage{amsmath}

\begin{document}
\title[Short Title]{Comment: Anomalous Anisotropic Light Scattering in Ge-Doped Silica Glass
}

\author{Saulius Juodkazis}
\thanks{Corresponding author}
\email{saulius@eco.tokushima-u.ac.jp}

\author{Shigeki Matsuo}

\author{Hiroaki Misawa}
\affiliation{Faculty of Engineering, The University of Tokushima, 2-1
Minamijyosanjima, Tokushima 770-8506, Japan}

\date{\today}



\maketitle


An \emph{anomalous anisotropic light scattering} which peaks in the
plane of light polarization, the ``propeller effect'' named by
authors, was observed in glass pumped by intense femtosecond laser
radiation~\cite{Kazansky}. A quiver movement of photoelectrons in the
light field was the key to an explanation of the effect. A blue
emission pattern of a propeller shape was attributed to a
defect-related photoluminescence (PL) and explained by its
scattering. The emission follows the angular distribution of the
photoelectrons given by $\frac{d\sigma}{d\Omega}\propto
(1+\cos^{2}\varphi)$, where the $\frac{d\sigma}{d\Omega}$ is the
differential scattering cross section of electrons, $\varphi$ is the
angle between the field amplitude vector $\mathbf{E}$ and the
electron momentum $\mathbf{k}_{e}$.

This is an interesting phenomenon relevant to a fast growing research
on the interaction on ultra-short pulses with materials, in
particular, with transparent solids. This is a field where some
long-lasting controversies over generation of white-light continuum
(super-continuum (SC)), self-focusing and dielectric breakdown (DB)
mechanisms should be resolved. By this comment we intend to draw
attention to the phenomenon of \emph{bremsstrahlung} (German word for
``braking radiation"), which escapes proper consideration and can
explain, for example, emission spectra of dielectric breakdown and
the above mentioned ``propeller effect''~\cite{Kazansky}. In our
judgement, the explanation of the ``propeller effect'' should take
into account \emph{bremsstrahlung}, transmission function of a
measurement setup, and optical aberrations.

\begin{figure}[t,b]
\includegraphics[width=7.75cm,draft=false]{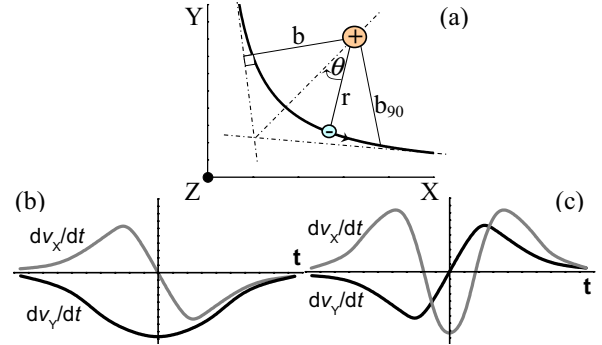}
\caption{(a) Geometry of electron-ion collision. Time dependence of
acceleration for straight-line, $b\ll b_{90}$, (b) and parabolic
$b\gg b_{90}$ (c) collisions ($b$ is the impact parameter; $b_{90}$
is the electron-ion distance at right angle scattering). At $t=0$ the
particles are at the closest. 
}\label{f-collision}
\end{figure}

Once the optical excitation generates free electrons, this not only
changes the dielectric function, which is usually considered in the
SC and light-induced DB mechanisms from the point of view of
nonlinear optics, but also inherently causes the light emission due
to the \emph{bremsstrahlung}. In a plasma state, after the dielectric
breakdown, when free charges are available, the radiated energy, $W$,
per unit angular frequency, $\omega$, in the nonrelativistic case is
given by:~\cite{Hutchinson}
\begin{equation}\label{brems}
\frac{dW}{d\omega} =
\frac{e^{2}}{6\pi^{2}\varepsilon_{0}c^{3}}\left|\int^{\infty}_{-\infty}\Dot{v}e^{i\omega
t}dt\right|^{2},
\end{equation}
where $\Dot{v}$ is the acceleration of an electron, which has a
charge of $e$, $c$ - speed of light, and $\varepsilon_{0}$ -
permittivity of vacuum. Spectral extent of \emph{bremsstrahlung}
depends on the type of electron-ion collisions
(Fig.~\ref{f-collision}) and can be defined as a reciprocal of the
collision time $\tau$. In the case of straight-line collisions
$\tau^{-1}\simeq v/(2b)$ while for the parabolic ones
$\tau^{-1}\simeq (v/(2b))(2b_{90}/b)^{2}$~\cite{Hutchinson}, $v$ is
the velocity of incoming electron. Emission spectrum of
\emph{bremsstrahlung} is continuous and spreads from X-rays-to-IR.
For an external observer (on z-axis) its spectral content depends on
the bandgap of the host material, defects' absorption, PL,
fractoluminescence, sonoluminescence, and a transmission function of
a measurement setup. Time resolved spectroscopy could, e.g.,
distinguish among those different contributions. For an isotropic
photoelectron distribution in plasma the eqn.~\ref{brems} defines an
emission spectrum integrated over all solid angles.

The actual polarization and angular dependence of the radiation of
quiver photoelectrons is defined by $\mathbf{R}\times
(\mathbf{R}\times \mathbf{\dot{v}})$~\cite{Hutchinson}, here
$\mathbf{R}$ is the vector from the charge to the field point (only a
far-field contribution is considered). Figure~\ref{f-collision}(b-c)
qualitatively shows the acceleration projections in the quiver plane
of photoelectron oscillating, e.g., along $\mathbf{E}_{x}$.
Obviously, there are considerable accelerations in the perpendicular
directions to the $\mathbf{E}_{x}$ for straight-line and parabolic
encounters. \emph{Bremsstrahlung} due to the $\mathbf{\dot{v}_{x}}$
will be projected into the $\mathbf{E}_{x}\propto\mathbf{R}_{Z}\times
(\mathbf{R}_{Z}\times \mathbf{\dot{v}}_{x})$, the quiver direction,
for an observer on z-axis ($\overrightarrow{z}$ is usually a
direction along the illumination/observation). Thus, the
xy-projection of \emph{bremsstrahlung} exactly follows the angular
distibution of photoelectrons and explains the ``propeller effect''.



\end{document}